\documentclass[12pt]{article}
\def\C{\mathbb C}

\def\E{\mathbb E}

\def\P{\mathbb P}
\def\R{\mathbb R}

\def\Z{\mathbb Z}

\def\wt{\widetilde}

\def\cH{{\mathcal H}}

\def\dist{{\rm{dist}}}
\def\diam{{\rm{diam}}}

\def\u0{{\underline 0}}

\def\uu{{\underline u}}

\def\ux{{\underline x}}
\def\uy{{\underline y}}

\def\wt{\widetilde}
\def\eps{{\epsilon}}
\def\om{{\omega}}
\def\Lam{{\Lambda}}

\def\pmn{\par\medskip\noindent}
\def\psn{\par\smallskip\noindent}
\def\z2{{\Z^2}}
\def\zp2{{\Z^2_{\geq}}}

\def\myset#1{{\left\{\,#1\,\right\}}}

\def\dist{{\,{\rm dist}}}

\usepackage{graphics}

\usepackage{latexsym}
\usepackage{amsfonts, amssymb}
\usepackage{amsmath}
\usepackage{amsthm}
\usepackage{psfrag}
\usepackage{pstricks,pst-node,pst-tree}
\usepackage{graphicx}
\usepackage{array}
\usepackage{rotating}

{\bf}{\it}
\newtheorem{Lem}{Lemma}[section]{\bf}{\it}
\newtheorem{Def}{Definition}[section]{\bf}{\it}
{\bf}{\it}

\begin{document}
\title{Wegner bounds for a two-particle\\ tight binding model}

\author{Victor Chulaevsky $^1$,
Yuri Suhov $^2$
}                     

\date{}

%
%
\maketitle
\pmn
\begin{center}
$^1$
D\'{e}partement de Math\'{e}matiques et Informatique, \\
Universit\'{e} de Reims, Moulin de la Housse, B.P. 1039, \\
51687 Reims Cedex 2, France \\
E-mail: victor.tchoulaevski@univ-reims.fr
\pmn
$^2$ Department of Pure Mathematics and Mathematical Statistics, \\
University of Cambridge, Wilberforce Road, \\
Cambidge CB3 0WB, UK\\
E-mail: Y.M.Suhov@statslab.cam.ac.uk
\end{center}
\newpage
\begin{abstract}
We consider a quantum two-particle system on a lattice $\Z^d$ with
interaction and in presence of an IID external potential.
We establish Wegner-typer estimates for such a model.
The main tool used is Stollmann's lemma.
\end{abstract}
\section{Introduction. The results}
\label{intro}

This paper considers a two-particle Anderson tight binding model
on lattice $\Z^d$ with  interaction. The Hamiltonian $H\left(=H^{(2)}_{U,V,g}
(\omega )\right)$ is a lattice Schr\"{o}dinger operator (LSO)
of the form $H^0+U+g(V_1+V_2)$ acting on functions
$\phi\in\ell_2(\Z^d\times\Z^d)$:
$$\begin{array}{cl}H\phi (\ux)&=H^0\phi (\ux)+
\left[\big(U+gV_1+gV_2\big)\phi\right](\ux )\\
\;&=\sum\limits_{\uy: \, \|\uy - \ux\|=1 } \, \phi(\uy)
+  \left(U(\ux)+g\sum_{j=1}^{2} V(x_j;\om) \right) \phi(\ux),\\
\;&\qquad\qquad\qquad \ux=(x_1,x_2),\;\uy =(y_1,y_2)\in\Z^d\times\Z^d.
\end{array}\eqno(1.1)$$
Here, $x_j=\big(x_j^{(1)},\ldots,x_j^{(d)}\big)$ and
$y_j=\big(y_j^{(1)},\ldots,y_j^{(d)}\big)$ stand for coordinate
vectors of the $j$-th particle in $\Z^d$,
$j=1,2$, and $\|\cdot\|$ is the sup-norm in $\R^d\times\R^d$:
$$\| x\| =\max_{j=1,2}\;\max_{i=1, \dots, d}\left|x_j^{(i)}\right|,
\;\;\ux = (x_1,x_2)\in\R^d\times\R^d.$$

Throughout this paper, the random external potential
$V(x;\omega )$, $x\in\Z^d$, is assumed to be
real IID, with the common distribution function $F$ on $\R$
satisfying the following condition:
\pmn
(I) {\sl $\forall$ $\epsilon >0$,}
$$s(\eps )\big(=s(F,\eps)\big)
:= \sup_{a\in \R} \,\, (F_V(a+\eps) - F_V(a) )  < \infty.
\eqno (1.2{\rm I})$$
\pmn
Finally, the interaction potential $U$ satisfies the following property:
\pmn
(II) {\sl $U$ is a bounded real function $[0,\infty )\to\R$ obeying}
$$U(\ux)=0,\;\hbox{ if }\;\|x_1-x_2\|>d.\eqno(1.2{\rm{II}})$$

The purpose of this paper is to establish the so-called
Wegner-type estimates for $H$. More precisely, these estimates
are produced for the eigen-values of a finite-volume approximation
$H_\Lam\left(=H^{(2)}_{\Lam,U,V,g}(\om )\right)$ (i.e., a $|\Lam |\times
\Lam |$ Hermitian matrix) acting on vectors in $\C^\Lam$:
$$\begin{array}{cl}H_\Lam\phi (\ux)&=H^0_\Lam\phi (\ux)+
\left[\big(U+gV_1+gV_2\big)_\Lam\phi\right](\ux )\\
\;&=\sum\limits_{\uy\in\Lam: \, \|\uy - \ux\|=1 } \, \phi(\uy)
+  \left(U(\ux)+g\sum_{j=1}^{2} V(x_j;\om) \right) \phi(\ux),\\
\;&\qquad\qquad\qquad \ux=(x_1,x_2),\;\uy =(y_1,y_2)\in\Lam\times\Lam.
\end{array}\eqno(1.3)$$
Here $\Lam\subset\Z^d\times\Z^d$ is a finite set of cardinality $|\Lam |$. For
definiteness, we will focus on the case where $\Lam$ is specified
as a $\Z^d\times\Z^d$ lattice cube written\\ as
the Cartesian product of two $\Z^d$ lattice cubes
centred at points\\ $u_1=\big(u_1^{1)},\ldots, u_1^{(d)}\big)\in\Z^d$
and $u_2=\big(u_2^{1)},\ldots, u_2^{(d)}\big)\in\Z^d$:
$$\left[\left(
{\operatornamewithlimits{\times}\limits_{i=1}^d}
\big[-L+u_1^{(i)},u_1^{(i)}+L\big]
\right)\times
\left({\operatornamewithlimits{\times}\limits_{i=1}^d}
\big[-L+u_2^{(i)},u_2^{(i)}+L\big]
\right)\right]\cap\big(\Z^d\times\Z^d\big).\eqno (1.4)$$
A set $\Lam$ of the form (1.4) will be called a box and
denoted by $\Lam_L(\uu )$,
$\uu=(u_1,u_2)\in\Z^d\times\Z^d$, while the $\Z^d$ lattice cubes
figuring
in the RHS (1.4) as the Cartesian factors
will be denoted by $\Pi_1\Lam_L(\uu)$ and $\Pi_2\Lam_L(\uu)$:
$$\Pi_j\Lam_L(\uu)=
\left({\operatornamewithlimits{\times}\limits_{i=1}^d}
\big[-L+u_j^{(i)},u_j^{(i)}+L\big]
\right)\cap\Z^d,\;\;j=1,2.\eqno (1.5)$$
We will also call cubes $\Pi_1\Lam_L(\uu)$ and $\Pi_2\Lam_L(\uu)$ the projections of
$\Lam_L(\uu )$. The cardinality of box $\Lam_L(\uu )$ is denoted by
$\left|\Lam_L(\uu )\right|$ and the cardinality of cube $\Pi_j\Lam_L(\uu)$
by $\left|\Pi_j\Lam_L(\uu )\right|$. \def\P{\mathbb P}
Symbol $\P$ will stand for
the probability distribution generated by random variables $V(x;\om )$,
$x\in\Z^d$. Symbol ${\mathfrak B}\left[\Lam_L(\uu )\right]$ is used for
the sigma-algebra generated by random variables
$$\om\mapsto V(x_1;\om )+V(x_2;\om ),\;\;\ux=(x_1,x_2)\in\Lam_L(\uu ).$$

The spectrum $\Sigma\left(H_{\Lam_L(\uu)}\right)$ of matrix
$H_{\Lam_L(uu)}$ is a random subset of $\R$ consisting of $\left|\Lam_L(\uu)
\right|$ points $\lambda^{(k)}_{\Lam_L(\uu )}$($=\lambda^{(k)}_{\Lam_L(\uu )}
(\om )$), $k=1,\ldots,\left|\Lam_L(\uu)\right|$ (random eigen-values in volume
$\Lam_L(\uu)$, measurable with respect to
${\mathfrak B}\left[\Lam_L(\uu )\right]$). Given a value $E\in\R$, we denote
$$\dist\big[\Sigma\left(H_{\Lam_L(\uu )}\right), E\big]
=\min\;\Big[\left|E-\lambda^{(k)}_{\Lam_L(\uu )}
\right|:\;k=1,\ldots ,\left|\Lam_L(\uu )\right|\Big].\eqno (1.6)$$
Our first result in this paper is the so-called single-volume Wegner
bound given in Theorem 1.

{\bf Theorem 1.} {\sl $\forall$ $E\in\R$, $L>1$, $\uu\in\Z^d\times\Z^d$
and $\epsilon >0$,}
$$\P\Big(\dist\left[\Sigma\left(H_{\Lam_L(\uu )}\right), E\right]
\leq \eps\Big)\leq
\left|\Lam_L(\uu )\right|\;\Big|\Pi_1\Lam_L(\uu )\cup
\Pi_2\Lam_L(\uu )\Big|\cdot s(2\eps).\eqno(1.7)$$

In Theorem 2 below we deal with a two-volume Wegner bound.
This bound assesses the probability
that the random spectra $\Sigma\left(H_{\Lam_L(\uu)}\right)$
and\\ $\Sigma\left(H_{\Lam_L(\uu')}\right)$ are close to each other,
for a pair of boxes $\Lam_L(\uu)$ and $\Lam_L(\uu')$
positioned away from each other, and conditional on
sigma-algebra ${\mathfrak B}\left[\Lam_L(\uu')\right]$.
More precisely, set:
$$\begin{array}{r}
\dist\left[\Sigma\left(H_{\Lam_L(\uu )}\right),
\Sigma\left(H_{\Lam_L(\uu')}\right)\right]
= \min\;\Big[\left|\lambda^{(k)}_{\Lam_L(\uu')}-\lambda^{(k')}_{\Lam_L(\uu')}
\right|:\qquad{}\\
k,k'=1,\ldots ,\left|\Lam_L(\uu )\right|\Big].\end{array}
\eqno (1.8)$$

  Our next result provides a probabilistic estimate on the distance between spectra in
two disjoint  boxes. An important feature of two-particle operators is that the potential
$W(u_1,u_2) = U(u_1,u_2) + g(V(u_1;\omega) + V(u_1;\omega))$ is a symmetric function of the pair
$(u_1,u_2)\in\Z^d$. Namely, let $S:\Z^2\times\Z^d$ be the following symmetry:
$$
S:\,(u_1,u_2) \mapsto (u_2,u_1).
$$
Then  $W(S(\ux)) \equiv W(\ux)$. As a consequence, spectra of operators $H_\Lam$ and
$H_{S(\Lam)}$ are identical.

{\bf Theorem 2.} {\sl $\forall$ $L>1$,
$\uu,\uu'\in\Z^d\times\Z^d$ with
$$
\min\, \{\|\uu -\uu'\|, \|S(\uu) -\uu'\| \}\geq 8L
\eqno(1.9)
$$
and $\epsilon >0$, at least one of the following inequalities holds: either}
$$\begin{array}{r}\P\Big(\dist\left[\Sigma\left(H_{\Lam_L(\uu )}\right),
\Sigma\left(H_{\Lam_L(\uu')}\right)\right]
\leq \eps\big|{\mathfrak B}\left[\Lam_L(\uu')\right]\Big)\qquad{}\qquad{}\\
\leq
\left|\Lam_L(\uu )\right|\;\left|\Lam_L(\uu')\right|\;
\Big|\Pi_1\Lam_L(\uu )\cup
\Pi_2\Lam_L(\uu )\Big|\cdot s(2\eps).\end{array}\eqno(1.10A)$$
{\sl or}
$$\begin{array}{r}\P\Big(\dist\left[\Sigma\left(H_{\Lam_L(\uu )}\right),
\Sigma\left(H_{\Lam_L(\uu')}\right)\right]
\leq \eps\big|{\mathfrak B}\left[\Lam_L(\uu)\right]\Big)\qquad{}\qquad{}\\
\leq
\left|\Lam_L(\uu )\right|\;\left|\Lam_L(\uu')\right|\;
\Big|\Pi_1\Lam_L(\uu' )\cup \Pi_2\Lam_L(\uu' )\Big|
    \cdot s(2\eps).\end{array}\eqno(1.10B)$$

The assertions of Theorems 1 and 2 are proved in the next section
of the paper, with the help of the
so-called Stollmann's lemma. They are useful in the
spectral analysis of $H$ and $H_{\Lam_L(\uu )}$. See
\cite{St2}. Note that in Theorem 1
we deal with the probability distribution $\P_{\Lam_L(\uu )}$
generated by the random variables
$$\om\mapsto V(x,\om ),\;x\in\Pi_1\Lam_L(\uu )\cup\Pi_2\Lam_L(\uu ),
\eqno (1.11)$$
whereas in Theorem 2 it is the conditional
probability distribution\\ $\P_{\Lam_L(\uu ),\Lam_L(\uu')}\big(\;\cdot\;\big|
{\mathfrak B}\left[\Lam_L(\uu')\right]\big)$ generated by
$$V(x,\;\cdot\;),\;x\in\Pi_1\Lam_L(\uu )\cup\Pi_2\Lam_L(\uu )
\cup\Pi_1\Lam_L(\uu')\cup\Pi_2\Lam_L(\uu')\eqno (1.12)$$
and conditioned relative to ${\mathfrak B}\left[\Lam_L(\uu')\right]$.

Throughout the paper, symbol $\qed$ is used to mark the end of
a proof.

\section{Stollmann's lemma. Proof of Theorems 1 and 2}

\subsection{Stollmann's lemma and its use}

For reader's convenience, we provide here the statement of Stollmann's lemma and its
proof; see Lemma 2.1 below. Cf. \cite{St1} and \cite{St2}, Lemma 2.3.1. Let $\Pi$ be a
non-empty finite set of cardinality $|\Pi |=p$. We assume that $\Pi $ is ordered and
identify it with the set $\{1, 2, \dots, p\}$. Consider the Euclidean space
$\R^\Pi$ of real dimension $p$, with standard basis
$({\mathbf e}_1, \dots, {\mathbf e}_p)$,
and its positive orthant
$$\R^\Pi _+ = \myset{{\mathbf v}=(q_1,\ldots ,q_p)\in\R^\Pi :
\, q_j\geq 0, \,\, j=1,\ldots, p }.$$
For a given probability measure $\mu$ on $\R$, denote by $\mu^\Pi$
the product measure $\mu
\times \dots \times \mu$ on $\R^\Pi $ and
by $\mu^{\Pi\setminus\{1\}}$ be the marginal product measure
induced by $\mu^\Pi$ on $\R^{\Pi\setminus\{1\}}$.
Next, $\forall$ $\epsilon >0$ set
$$s(\mu, \eps) = \sup_{a\in\R} \,\, \int_{a}^{a+\eps} d\mu(t)
\eqno (2.1)$$
and assume that $s(\mu,\eps)<\infty$.

\begin{Def} A function $\Phi:\, \R^\Pi \to\R$
is called diagonally-monotone (DM) if it satisfies the following
conditions:
\psn
{\rm (i)} $\forall$ ${\mathbf r}\in\R^p_+$ and any ${\mathbf v}\in \R^p$,
$$
\Phi({\mathbf v}+{\mathbf r})\geq \Phi({\mathbf v});\eqno (2.2)$$
\psn
{\rm (ii)} moreover, with vector ${\mathbf e}={\mathbf e}_1
+ \ldots + {\mathbf e}_p\in \R^p$, $\forall$ ${\mathbf v}\in\R^p$
and $t>0$
$$\Phi({\mathbf v} + t{\mathbf e}) - \Phi({\mathbf v}) \geq t.
\eqno (2.3)$$
\end{Def}

\begin{Lem}\label{Stoll} Suppose function $\Phi:\, \R^\Pi  \to \R$ is
DM. Then $\forall$ $\eps >0$ and any
open interval $I\subset \R$ of length $\eps$,
$$\mu^\Pi\myset{ {\mathbf v}:\, \Phi({\mathbf v}) \in I }
\leq p \cdot s(\mu,\eps ).\eqno (2.4)$$
\end{Lem}
\proof
Let $I=(a,b)$, $b-a=\eps$, and consider the set
$$A = \myset{ {\mathbf v}:\, \Phi({\mathbf v}) \leq a  }.$$
Furthermore, define recursively sets $A^\eps_j$, $j=0, \dots, p$, by setting
$$A^\eps_0 = A, \; A^\eps_j = A^\eps_{j-1} + [0,\eps]{\mathbf e}_j :=
\, \myset{ {\mathbf v}+t{\mathbf e}_j:\,
{\mathbf v}\in A^\eps_{j-1}, \,\, t\in[0,\eps]  }.$$
Obviously, the sequence of sets $A^\eps_j$, $j=1, 2, ...$, is
increasing with $j$. The DM property implies that
$\myset{ {\mathbf v}:\, \Phi({\mathbf v}) < b  } \subset A^\eps_p$.
Indeed, if $\Phi({\mathbf v}) < b$, then for the vector ${\mathbf v}':
={\mathbf v}- \eps\cdot{\mathbf e}$ we have by property (ii):
$$
\Phi({\mathbf v}') \leq \Phi({\mathbf v}' + \eps \cdot{\mathbf e}) - \eps
= \Phi({\mathbf v}) - \eps \leq b - \eps \leq a,
$$
meaning that ${\mathbf v}'\in \myset{ \Phi \leq a} = A$ and, therefore,
${\mathbf v} = {\mathbf v}' + \eps\cdot{\mathbf e}\in A^\eps_p$.
We conclude that
$$\myset{ {\mathbf v}:\, \Phi({\mathbf v}) \in I}
= \myset{ {\mathbf v}:\, \Phi({\mathbf v}) < b  }
\setminus \myset{ {\mathbf v}:\, \Phi({\mathbf v}) \leq a  }
\subset A^\eps_m \setminus A.$$
Moreover, the probability $\mu^\Pi\myset{ {\mathbf v}:\,
\Phi({\mathbf v}) \in I }$ is
$$\leq
\mu^\Pi\left( A^\eps_p \setminus A \right)
= \mu^\Pi\left( \bigcup_{j=1}^p\left(  A^\eps_j \setminus A^\eps_{j-1} \right)
\right)
\leq \sum_{j=1}^p \mu^\Pi\left( A^\eps_j \setminus A^\eps_{j-1} \right).
$$\def\wt{\widetilde}

For ${\wt{\mathbf v}}\in\R^{\Pi\setminus\{1\}}$, set
$I_1({\wt{\mathbf v}}) = \myset{ q_1\in\R:\, (q_1,{\wt{\mathbf v}})
\in A^\eps_1\setminus A}$. Then,
by definition of set $A^\eps_1$, set $I_1({\wt{\mathbf v}})$ is an interval
of length $\leq \eps$. Thus,
$$\mu^\Pi( A^\eps_1 \setminus A) = \int d\mu^{\Pi\setminus\{1\}}
({\wt{\mathbf v}}) \, \int_{I_1({\wt{\mathbf v}})} d\mu(q_1)
\leq s(\mu, \eps).$$
Similarly, for $j=2, \dots, p$ we obtain
$\mu^\Pi( A^\eps_j \setminus A^\eps_{j-1}) \leq s(\mu,\eps)$,
which yields that
$$\mu^\Pi\myset{ {\mathbf v}:\, \Phi({\mathbf v}) \in I } \leq
\sum_{j=1}^p \mu^\Pi( A^\eps_j \setminus A^\eps_{j-1})
\leq p \cdot s(\mu, \eps). \;\;\;\qed
$$

In our situation, it is also convenient to introduce the notion of
a DM operator family.

\begin{Def} Let $\cH$ be a
Hilbert space of a finite dimension $m$. A family of Hermitian operators
$B({\mathbf v}):\cH \to \cH$, ${\mathbf v}\in\R^\Pi$,
is called DM if, $\forall$
$f\in\cH$
$$(B({\mathbf v}+t\cdot{\mathbf e})f, f)
- (B({\mathbf v})f,f) \geq t\cdot \|f\|^2.\eqno (2.5)$$
That is, $\forall$
$f\in\cH$  with $\|f\|=1$, the function $\Phi_f: \R^\Pi  \to \R$ defined by
$\Phi_f({\mathbf v}) = (B({\mathbf v})f, f)$ is DM.
\end{Def}\pmn

{\bf Remark 2.1.} Suppose that $B({\mathbf v})$, ${\mathbf v}\in\R^\Pi$,
is a DM operator family in $\cH$. Let $E^{(1)}_{B({\mathbf v})}\leq\ldots\leq
E^{(m)}_{B({\mathbf v})}$ be the eigen-values of $B({\mathbf v})$.
Then, by virtue of the variational principle,
$\forall$ $k=1,\ldots,m$,
${\mathbf v}\mapsto E^{(k)}_{B({\mathbf v})}$ is a DM function.
\psn

{\bf Remark 2.2.}  If  $B({\mathbf v}),\,{\mathbf v}\in\R^\Pi$, is a
DM operator family
in $\cH$, and $K:\cH\to\cH$ is an arbitrary Hermitian operator, then
the family $K+ B({\mathbf v})$ is also DM.

\subsection{Proof of Theorems 1 and 2}

{\it Proof of Theorem} 1. The proof is a straightforward application of
Lemma 2.1 and Remarks 2.1 and 2.2. Cf. the proof of Theorems 2.3.2 and 2.3.3
in \cite{CS1}. For a single-particle tight binding model, similar results
are presented in \cite{Ch1}.
In our situation, set $\Pi$ is identified as the union
$\Pi_1\Lam_L(\uu )\cup\Pi_2\Lam_L(\uu )$, with
$p=\left|\Pi_1\Lam_L(\uu )\cup\Pi_2\Lam_L(\uu )\right|$. Vector ${\mathbf v}$
is identified with a collection $\{V(x,\om ),\;x\in
\Pi_1\Lam_L(\uu )\cup\Pi_2\Lam_L(\uu )\}$ of sample values of the
external potential; to stress this fact we will write
$${\mathbf v}\sim
\{V(x,\om ),\;x\in\Pi_1\Lam_L(\uu )\cup\Pi_2\Lam_L(\uu )\}.\eqno (2.6)$$
Next, probability measure $\mu$ represents the distribution
of a single value, say $V(0,\;\cdot\;)$, and product-measure
$\mu^\Pi$ is identified as $\P_{\Lam_L(\uu )}$. Further, the Hilbert
space $\cH$ in Remarks 2.1 and 2.2 is $\C^{\Lam_L(\uu )}$, of dimension
$m=\left|\Lam_L(\uu )\right|$, in which
the action of matrix $H_{\Lam_L(\uu )}$ is considered. Given
$\ux=(x_1,x_2)\in\Lam_L(\uu )$, we can write
$$g\big[V(x_1,\om )+V(x_2,\om )\big]
=\sum_{y\in\Pi_1\Lam_L(\uu )\cup\Pi_2\Lam_L(\uu )}c(\ux,y)V(y,\om )$$
where $c(\ux,y)=\begin{cases}1,&y=x_1\;\hbox{ or }\;x_2,\\
0,&y\neq x_1,x_2.\end{cases}$ This implies that, with
identification (2.6), operators Hermitian $B({\mathbf v})$
form a DM family. Here $B({\mathbf v})$ is the multiplication operator
$$B({\mathbf v})\phi (\ux)=g\big[V(x_1,\om )+V(x_2,\om )\big]\phi (x),
\;\;\ux\in\Lam_L(\uu ),\;\phi\in\C^{\Lam_L(\uu )}\eqno (2.7)$$
Then we use Remark 2.2, with $K=H^0_{\Lam_L(\uu )}+U$ (cf. (1.3)),
and obtain that $H_{\Lam_L(\uu )}=K+B({\mathbf v})$ is a DM family.
Next, owing to Remark 2.1, each eigen-value $\lambda^{(k)}_{\Lam_L(\uu )}$,
$k=1,\ldots,\left|\Lam_L(\uu )\right|$, is a DM function of the
sample collection $\{V(x,\om ),\;x\in\Pi_1\Lam_L(\uu )\cup\Pi_2\Lam_L(\uu )\}$.
Hence, by Lemma 2.1, $\forall$ $k=1,\ldots ,\left|\Lam_L(\uu )\right|\Big]$,
$$\P\Big(\left|E-\lambda^{(k)}_{\Lam_L(\uu )}
\right|\leq\eps\Big)\leq \left|\Pi_1\Lam_L(\uu )\cup\Pi_2\Lam_L(\uu )\right|
s(F,2\eps),\eqno (2.8)$$
The final remark is, that the probability in the LHS of Eqn (1.7) is
$\leq$ the RHS of Eqn (2.8) times $\left|\Lam_L(\uu )\right|$.
$\quad\qed$
\pmn

  We will need the following elementary geometrical statement.

\begin{Lem}\label{Geom} Consider two boxes $\Lam_L(\uu )$ and $\Lam_L(\uu')$
and suppose that
$$
\min(\|\uu -\uu'\|, \|S(\uu) -\uu'\|) \geq 8L \eqno (2.9)
$$
Then there are two possibilities (which in general do not exclude each other):

(i) $\Lam_L(\uu )$ and $\Lam_L(\uu')$
are `completely separated', when
$$\left(\Pi_1\Lam_L(\uu )\cup\Pi_2\Lam_L(\uu )\right)\cap
\left(\Pi_1\Lam_L(\uu')\cup\Pi_2\Lam_L(\uu')\right)=\emptyset.
\eqno (2.10)$$

(ii) $\Lam_L(\uu )$ and $\Lam_L(\uu')$
are `partially separated'. In this case one (or more) of the four possibilities can
occur:
$$\begin{array}{lll}{\rm{(A)}}
& \Pi_1\Lam_L(\uu )\cap\left[\Pi_2\Lam_L(\uu )\cup
\Pi\Lam_L(\uu')\right]&=\emptyset,\\
{\rm{(B)}}& \Pi_2\Lam_L(\uu )\cap\left[\Pi_1\Lam_L(\uu )\cup
\Pi\Lam_L(\uu')\right]&=\emptyset,\\
{\rm{(C)}}& \Pi_1\Lam_L(\uu')\cap\left[\Pi\Lam_L(\uu )\cup
\Pi_2\Lam_L(\uu')\right]&=\emptyset,\\
{\rm{(D)}}& \Pi_2\Lam_L(\uu')\cap\left[\Pi\Lam_L(\uu )\cup
\Pi_1\Lam_L(\uu')\right]&=\emptyset,\end{array}\eqno (2.11)$$
where
$$\Pi\Lam_L(\uu')=\Pi_1\Lam_L(\uu')\cup\Pi_2\Lam_L(\uu'),\;\;
\Pi\Lam_L(\uu )=\Pi_1\Lam_L(\uu )\cup\Pi_2\Lam_L(\uu ).\eqno (2.12)$$
\end{Lem}

  Pictorially,  case (ii) is where one of the cubes
$\Pi_j\Lam_L(\uu )$, $\Pi_j\Lam_L(\uu')$,
$j=1,2$,
is disjoint from the union of the rest of the projections of $\Lam_L(\uu )$ and
$\Lam_L(\uu')$.

\pmn
{\it Proof of Theorem} 2. Owing to Lemma 2.2, boxes  $\Lam_L(\uu )$ and
$\Lam_L(\uu')$ satisfy either (i) or (ii), i.e. they are either completely or partially
separated.  We note that the use of the max-norm $\|\;\|$ is convenient here as it leads
to the constant $8$ (equal to $2$ times $4$, the number of projections $\Pi_j\Lam_L(\uu
)$ and $\Pi_j\Lam_L(\uu')$, $j=1,2$) which does not depend on the dimension $d$.

Passing to the proof of Theorem 2 proper, note first that, under the conditional
probability distribution in Eqn (1.10A), the eigen-values
$\lambda^{(k')}_{\Lam_L(\uu ')}$,
$k'=1,\ldots ,\left|\Lam_L(\uu ')\right|$, forming the set
$\Sigma\left(H_{\Lam_L(\uu')}\right)$ are non-random. The same is true, of course, for
 the eigen-values
$\lambda^{(k')}_{\Lam_L(\uu ')}$,
$k'=1,\ldots ,\left|\Lam_L(\uu ')\right|$ in Eqn (1.10B).
Now, by virtue of (2.10), (2.11), the boxes
$\Lam_L(\uu )$ and $\Lam_L(\uu')$ are either completely or partially
separated. In the former case, the conditional probability distribution
in (1.9) is reduced to the probability measure $\P_{\Lam_L(\uu )}$.
Then, as in the proof of Theorem 1 (cf. Eqn (2.8)), $\forall$
$k=1,\ldots ,\left|\Lam_L(\uu ')\right|$, for the (random) eigen-value
$\lambda^{(k)}_{\Lam_L(\uu )}\in\Sigma\left(H_{\Lam_L(\uu )}\right)$
the following bound holds true:
$$\P\Big(\left|\lambda^{(k)}_{\Lam_L(\uu ')}-
\lambda^{(k')}_{\Lam_L(\uu ')}\right|
\leq \eps\big|{\mathfrak B}\left[\Lam_L(\uu')\right]\Big)
\leq \left|\Pi_1\Lam_L(\uu )\cup\Pi_2\Lam_L(\uu )\right|
s(F,2\eps),\eqno (2.13)$$ implying bound (1.10A).

Now assume that $\Lam_L(\uu )$ and $\Lam_L(\uu')$ are partially separated. For example,
assume case A where $\Pi_1\Lam_L(\uu )$, is disjoint from the union of the rest of the
projections of $\Lam_L(\uu )$ and $\Lam_L(\uu')$:
$$\Pi_1\Lam_L(\uu )\cap\left[\Pi_2\Lam_L(\uu )\cup\Pi\Lam_L(\uu'))\right]
=\emptyset.\eqno (2.14)$$

We then write the probability in the LHS
of (2.13) as the conditional expectation
$$\begin{array}{l}\P\Big(\left|\lambda^{(k)}_{\Lam_L(\uu ')}-
\lambda^{(k')}_{\Lam_L(\uu ')}\right|
\leq \eps\big|{\mathfrak B}\left[\Lam_L(\uu')\right]\Big)\\
=\E\left[\P\Big(\left|\lambda^{(k)}_{\Lam_L(\uu ')}-
\lambda^{(k')}_{\Lam_L(\uu ')}\right|\leq \eps\big|
{\mathfrak C}\big[\Pi_2\Lam_L(\uu )\cup
\Pi\Lam_L(\uu')\big]\Big)
\Big|{\mathfrak B}\left[\Lam_L(\uu')\right]\right].\end{array}
\eqno (2.15)$$
Here ${\mathfrak C}\big[\Pi_2\Lam_L(\uu )\cup
\Pi\Lam_L(\uu')\big]$ is the sigma-algebra generated by the
random variables
$$\om\mapsto V(x,\om ),\;\;x\in\Pi_2\Lam_L(\uu )\cup
\Pi\Lam_L(\uu');$$
owing to (2.14) it is independent of the sigma-algebra
${\mathfrak C}\left[\Pi_1\Lam_L(\uu )
\right]$ generated by the random variables
$$\om\mapsto V(x,\om ),\;\;x\in\Pi_1\Lam_L(\uu ).$$

We see that the argument used in the proof of Theorem 1 is still applicable,
if we replace the product-measure $\P_{\Lam_L(\uu )}$ by its
restriction
to ${\mathfrak C}\left[\Pi_1\Lam_L(\uu )\right]$ (which
again can be taken as a product-measure $\mu^\Pi$ from Lemma 2.1, with $p=
\left|\Pi_1\Lam_L(\uu )\right|$). This allows us to write
$$\P\Big(\left|\lambda^{(k)}_{\Lam_L(\uu ')}-
\lambda^{(k')}_{\Lam_L(\uu ')}\right|\leq \eps\big|
{\mathfrak C}\big[\Pi_2\Lam_L(\uu )\cup
\Pi\Lam_L(\uu')\big]\Big)\leq\left|\Pi_1\Lam_L(\uu )\right|
s(F,2\eps)\eqno (2.16)$$ and deduce a similar bound for the the conditional probability
in the LHS of (2.15). Inequality (1.10A) is then derived in the standard manner.

If, instead of (2.14), we have one of the other disjointedness relations (B)-(D) in Eqn
(2.11) then the argument is conducted in a similar fashion. Naturally, in the case (B) we
still prove (1.10A), while in the cases (C) and (D) we prove (1.10B).

 This concludes the proof of Theorem 2.

\subsection{Proof of Lemma 2.2}

  Recall that we have two boxes, $\Lam_L(\uu )$ and $\Lam_L(\uu' )$, satisfying the
condition (2.9):
$$
\min(\|\uu -\uu'\|, \|S(\uu) -\uu'\|) \geq 8L.
$$
Notice that this can be viewed as lower bound for the distance in the factor space
$\Z^d\times\Z^d / S$; recall that $S(u_1,u_2)=(u_2,u_1)$.

Since $\diam\, \Lam_L(\uu ) = \diam\, \Lam_L(\uu' )=2L$, this implies that the union of
the four coordinate projections,
$$
\Pi_1\Lam_L(\uu ), \Pi_2\Lam_L(\uu ), \Pi_1\Lam_L(\uu'), \Pi_2\Lam_L(\uu')
$$
cannot be connected. Therefore, it can be decomposed into two or more connected
components. Cases $(A)$, $(B)$, $(C)$ and $(D)$ in the statement of Lemma 2.2 correspond
to the situation where one of these coordinate projections is disjoint with the three
remaining projections. So, it suffices to analyse the case where each connected component
of the union
$$
\Pi_1\Lam_L(\uu ) \cup \Pi_2\Lam_L(\uu ) \cup \Pi_1\Lam_L(\uu') \cup \Pi_2\Lam_L(\uu')
\eqno(2.17)
$$
contains exactly two coordinate projections. Furthermore, it suffices to show that the
only possible case is (2.10)  where
$$
\left(\Pi_1\Lam_L(\uu )\cup\Pi_2\Lam_L(\uu )\right)\cap
\left(\Pi_1\Lam_L(\uu')\cup\Pi_2\Lam_L(\uu')\right)=\emptyset.
$$
To do so, we have to exclude two remaining cases, namely,
$$
\begin{cases}
    \left(\Pi_1\Lam_L(\uu )\cup\Pi_1\Lam_L(\uu' )\right)\cap
    \left(\Pi_2\Lam_L(\uu)\cup\Pi_2\Lam_L(\uu')\right)=\emptyset \\
    \Pi_1\Lam_L(\uu ) \cap \Pi_1\Lam_L(\uu' ) \neq \emptyset \\
    \Pi_2\Lam_L(\uu)\cup\Pi_2\Lam_L(\uu') \neq \emptyset
\end{cases}
\eqno(2.18)
$$
and
$$
\begin{cases}
    \left(\Pi_1\Lam_L(\uu )\cup\Pi_2\Lam_L(\uu' )\right)\cap
    \left(\Pi_1\Lam_L(\uu')\cup\Pi_2\Lam_L(\uu)\right)=\emptyset \\
    \Pi_1\Lam_L(\uu )\cup\Pi_2\Lam_L(\uu' ) \neq \emptyset \\
    \Pi_1\Lam_L(\uu')\cup\Pi_2\Lam_L(\uu) \neq \emptyset
\end{cases}
\eqno(2.19)
$$
First, observe that (2.18) contradicts the assumption that
$\Lam_L(\uu)$ and $\Lam_L(\uu')$ are disjoint (and even distant). Indeed, in such a case,
there exist lattice points
$$
v_1\in \Pi_1\Lam_L(\uu )\cup\Pi_2\Lam_L(\uu' ),
\; v_2\in\Pi_2\Lam_L(\uu)\cup\Pi_2\Lam_L(\uu'),
$$
so that
$$
\begin{array}{cl}
\exists\, (v_1,v_2) \in
\left[ \Pi_1\Lam_L(\uu ) \times \Pi_2\Lam_L(\uu ) \right]
\cap \left[\Pi_1\Lam_L(\uu' ) \times \Pi_2\Lam_L(\uu' ) \right]\\
= \Lam_L(\uu) \cap \Lam_L(\uu') = \emptyset,
\end{array}
$$
which is impossible.

   The case (2.19) can be reduced to (2.18), by the symmetry $S$. Namely, let
$\uu'' = S(\uu')$, then
$$
\Pi_1 \Lam_L(\uu'') = \Pi_2 \Lam_L(\uu'),
\, \Pi_2 \Lam_L(\uu'') = \Pi_1 \Lam_L(\uu').
$$
Now (2.19) reads as follows  in terms of boxes $\Lam_L(\uu)$, $\Lam_L(\uu')$:
$$
\begin{cases}
    \left(\Pi_1\Lam_L(\uu )\cup\Pi_1\Lam_L(\uu'' )\right)\cap
    \left(\Pi_2\Lam_L(\uu'')\cup\Pi_2\Lam_L(\uu)\right)=\emptyset \\
    \Pi_1\Lam_L(\uu )\cup\Pi_1\Lam_L(\uu'' ) \neq \emptyset \\
    \Pi_2\Lam_L(\uu'')\cup\Pi_2\Lam_L(\uu) \neq \emptyset.
\end{cases}
\eqno(2.20)
$$
The same argument as above shows then that
$\Lam_L(\uu) \cap \Lam_L(\uu'') \neq \emptyset$, which is impossible, since
$$
\dist( \uu, S(\uu') ) > 8L.
$$
This completes the proof.
\qed

\section{Concluding remarks}

{\bf Acknowledgments.}
VC thanks The Isaac Newton Institute and Department
of Pure Mathematics and Mathematical Statistics, University of Cambridge,
for the hospitality during visits in 2003, 2004 and 2007. YS thanks
D\'{e}partement de Mathematique, Universit\'{e} de Reims
Champagne--Ardenne, for the hospitality during visits in 2003 and 2006,
in particular, for the Visiting Professorship in the Spring of 2003.
YS thanks IHES,
Bures-sur-Yvette, France, for the hospitality during numerous visits
in 2003--2007. YS thanks School of Theoretical Physics, Dublin Institute
for Advanced Studies, for the hospitality during regular visits in
2003--2007. YS thanks Department of Mathematics, Penn State University,
for the hospitality during Visting Professorship in the Spring,
2004. YS thanks Department of Mathematics, University of California, Davis,
for the hospitality during Visiting Professroship in the Fall of 2005.
YS acknowledges the support provided by the ESF Research Programme RDSES
towards research trips in 2003--2006.

%
%

\end{document}